\documentclass{jltp}
\usepackage{graphicx,color,latexsym} 
\title{Analysis of possible field-induced superconducti- vity 
in anthracene, other
polyacenes, and C$_{60}$}
\author{S.-L.\ Drechsler, H. Rosner$^*$, G.\ Paasch, J.\ M\'alek$^+$, 
H.\ Eschrig}
\address{Leibniz-Institut f\"ur Festk\"orper- und Werkstofforschung Dresden, 
P.O. Box 270116, \\
D-01171 Dresden, Germany\\
$^*$Department of Physics, University of California, 
Davis, CA, 95616, USA\\
$^+$Institute of Physics, ASCR, Na Slovance 2, CZ-18221 Praha 8, 
Czech Republic}
\runninghead{S.-L.\ Drechsler, H.\ Rosner, G.\ Paasch, et al.}
{ Analysis of possible field-induced 
superconductivity}
\begin{document}
\maketitle

\begin{abstract}
We consider electronic structure and superconductivity aspects
in field-doped polyacenes (PA) and C$_{60}$. 
Within a modified Thomas-Fermi approach   for typical 
experimental values of the surface 
charge density the 
injected charge is confined to a monolayer.  
The electron-phonon coupling constant 
for internal modes
$\lambda_{int}$ is estimated using the work of Devos et al.\
({\it Phys.\ Rev.\ B} {\bf 58}, 8236 (1998)) 
and the density of states 
estimated from  a 2D-one-band 
model derived from a full potential LDA 
band structure calculation for bulk anthracene. 
The large differences in the reported $T_c$-values  for 
 PA and C$_{60}$ are ascribed to enhanced empirical  
Coulomb pseudopotentials $\mu^*$ for PA.
\end{abstract}

Recently   
field-induced superconductivity (FISC) has been reported
for  field effect transistors (FET) based on various   organic 
systems.\cite{schoen00a,schoen00c,batlogg01} Unfortunately, 
its very existence and the possibility to produce the necessary 
extremly high electric fields in FETs have not been verified so far by 
other groups.\cite{science}
 Anyhow, an 
electron-phonon (el-ph) mechanism 
is considered  
to be a likely candidate\cite{schoen00a,schoen00c}, 
although, details of the FISC  are still unknown. 
Due to the closely related physics of bulk 
electron $n$-doped
A$_3$C$_{60}$
(A=K, Rb) the main contributions from high-frequency internal molecular
optical phonons  
$\hbar \omega_{ph} \sim$ 100 meV
could be expected.\cite{gunnarsson92} 
There is a sizable corresponding 
el-ph coupling 
in 
 aromatic molecules  in general.\cite{devos98}
 Its
 strength 
  increases
with decreasing number of  $\pi$-electrons $N_\pi$, 
 in line with the  
 $T_c$ values (2, 2.7, and 4 K) reported for 
 $n$-doped penta-, tetra-, and anthracene, respectively).
 Anyhow, with respect to PA 
 the reason for 
 the much higher $T_c \approx $  
 11 (18 - 26) K, 52 (80 -114) K for 
 $n$- and $h$-doped C$_{60}$, respectively, 
 (values in brackets for haloforms CHCl(Br)$_3$ 
  intercalated\cite{duennebier},
 additional coupling to internal modes of these haloforms
 has been proposed\cite{bill})
  and the role of 
 Coulomb interaction
 remain to be understood.
According to qualitative
 arguments \cite{schoen00a} the FI charge is
 accumulated $\approx$ only
within 
a {\it mono}layer at the surface of the organic semiconductor.
 \noindent
\begin{figure}[t]
\begin{center}
        \includegraphics[scale=0.5]{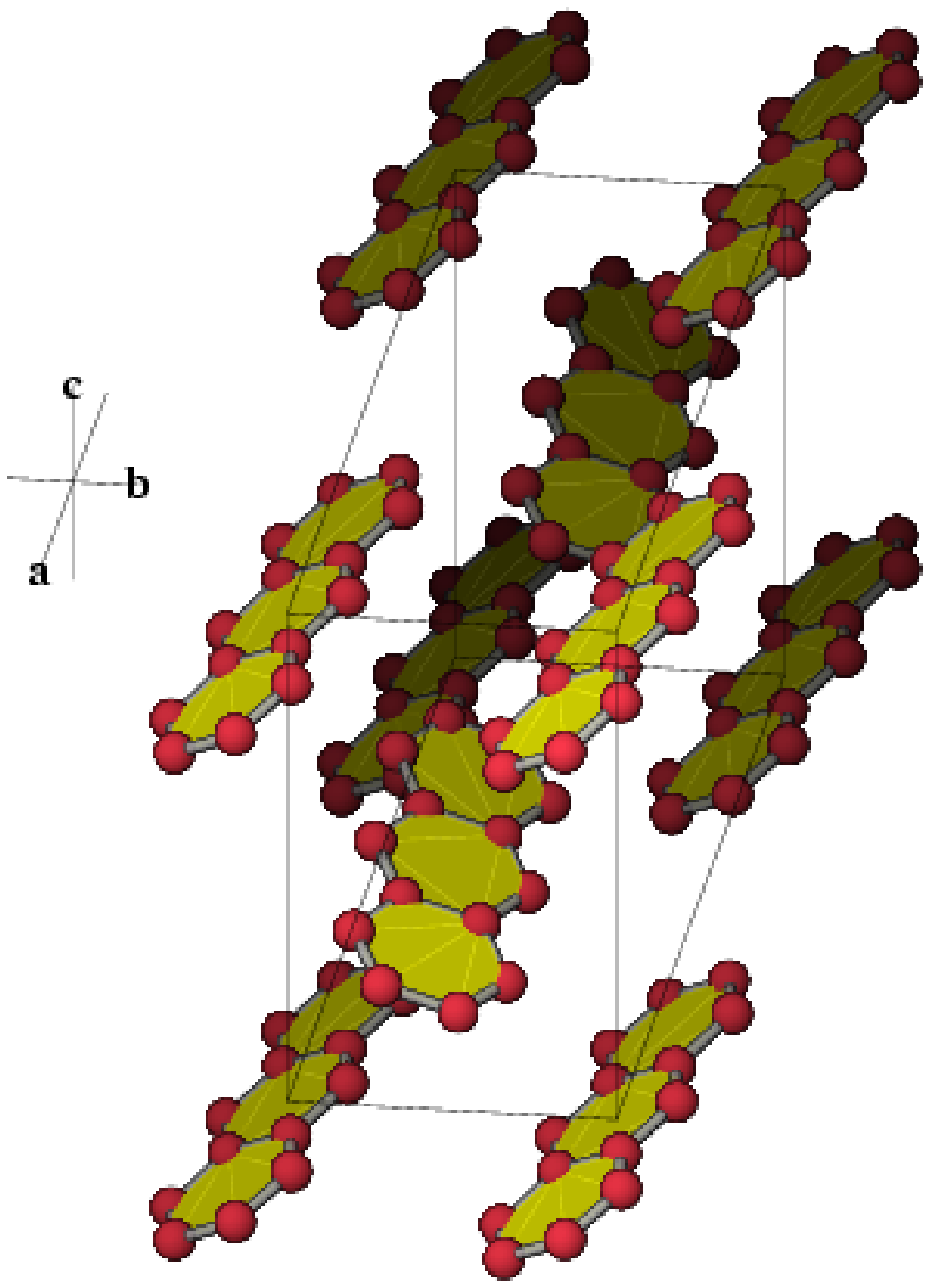}
\hspace{0.5cm}
\vspace{0.5cm}
        \includegraphics[scale=0.4]{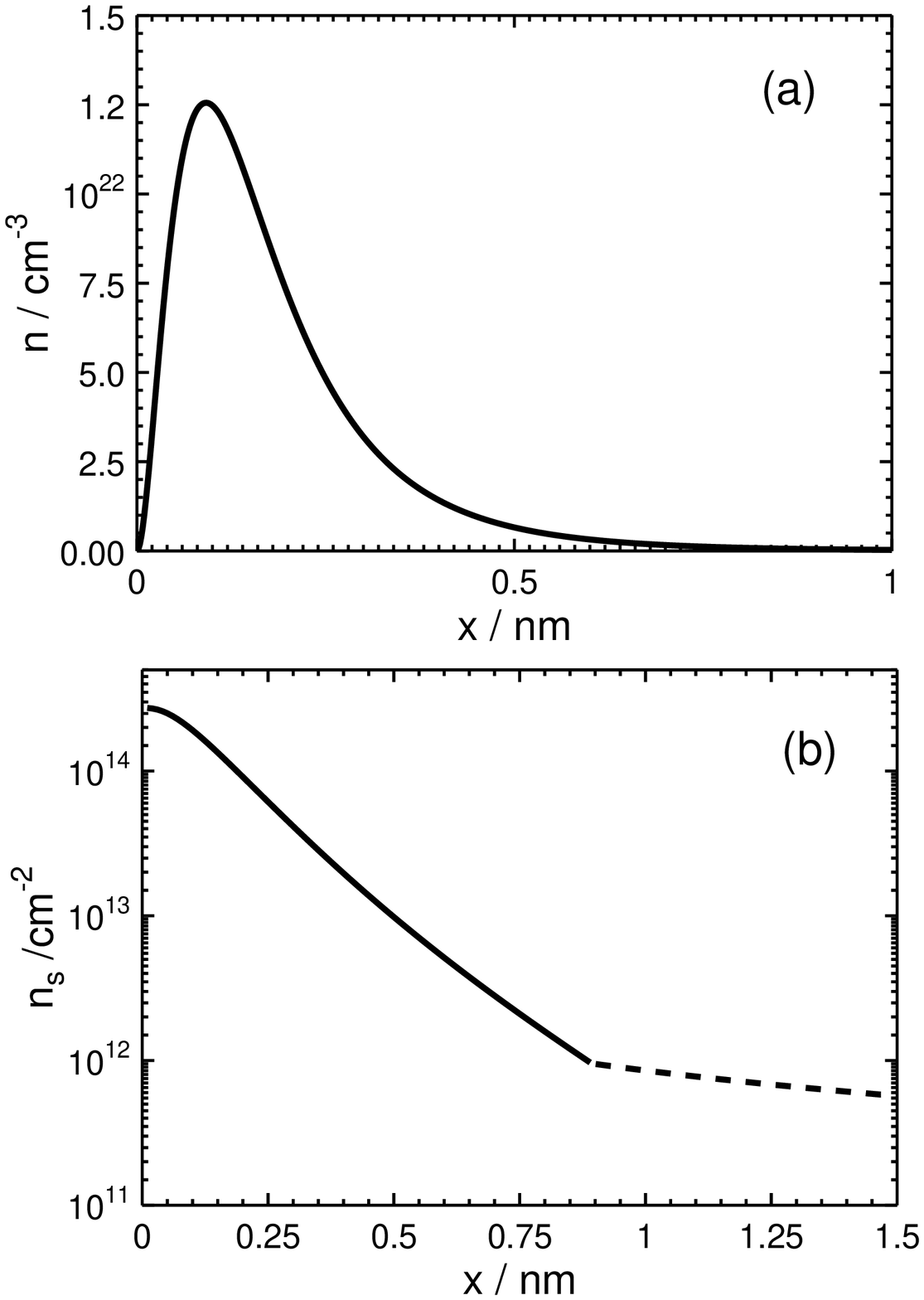}\\
        \end{center}
\vspace{-0.2cm}
\caption{ (Left panel) Crystal structure of anthracene. The FI charges  
in a FET move $\|$ to the surface  within the 
$bc$-plane. 
 (Right panel) Calculated volume density $n$ (a) and   area charge 
density $n_s$ (b) 
{\it vs.\ }  distance $x$ 
to the oxide within a FET. In (b): full line - degenerate 
statistics approximation with quantum
confinement, dashed line - nondegenerate statistics.}
\vspace{-0.0cm}
 \label{ancryst}
 \end{figure}
 Since it is important to know whether the 
 FISC is confined 
to such an effective 2D subsystem, 
we check first that assumption. 
 Secondly,  
we report local density approximation (LDA) 
  band structure 
calculations for the prototypical   undoped   bulk anthracene.
Consequences for novel many-body physics and possible FISC are 
briefly discussed.

In the FET the injected charge can be adjusted by the gate bias. 
 Concentrations $n_s >$ 
 10$^{14}$ electrons/cm$^2$ can be 
 accumulated at the surface of 
 organic semiconductors. 
 Such densities correspond to 
 about one (three) electron(s) per PA (C$_{60}$)
 molecule, if one assumes
that only
the topmost molecular layer 
takes part in the conduction and in the FISC
 (see Fig.\ 1). 
 We calculated   
the
concentration  and the
   field profiles
  within a quasiclassical 
  approximation for
$n$-doped C$_{60}$. The undoped C$_{60}$ 
is treated as an intrinsic semiconductor
    (gap 2eV). Assuming for
    the
effective density of states 1$\times 10^{20}$cm$^{-3}$ the narrow band width 
(large
effective mass)
    is taken into account\cite{paasch98a}   
    and a dielectric constant
    $\epsilon = 3 - 4$  is
used.\cite{knupfer} For $E_F$ still in the gap, the carrier
concentration is given by non-degenerate statistics and otherwise by Fermi
statistics. Here,  we additionally take into
account   that the charge density of the inversion (accumulation) 
layer practically
vanishes at the oxide interface  due to the large potential barrier 
caused by the big oxide gap $\sim$ 8 - 10 eV.
This quantum confinement effect can be incorporated by using
the modified Thomas-Fermi approximation.\cite{paasch81} An example
 is shown in Fig.\ 1 
for a total area density\cite{schoen00a}
$n_s=2.7 \times 10^{14}$cm$^{-2}$ (corresponding to a surface electric field of
1.6$\times 10^{8}$V/cm). Fig.\ 1a shows the density as a function of the
 distance $x$ to
the oxide on a linear scale and Fig.\ 1b the area 
density (from $x$ to
infinity). 
\begin{figure}[t]
\begin{center}
\includegraphics[scale=0.34,angle=-90]{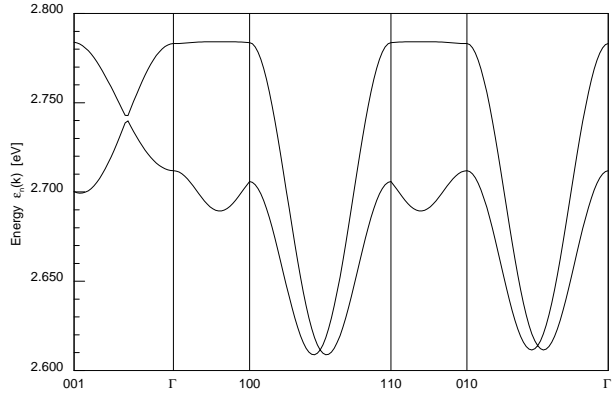}
\end{center}
\vspace{0.cm}
\caption{ The LDA-conduction band of bulk undoped anthracene.
Notice its
anisotropic quasi-2D like nature: a 
 strong unusual 
 dispersion along the $b$-axis, a reduced 
 one along $c$,  
 and a very weak one along
the $a$-axis.}
\end{figure}
\vspace{-0.0cm}
The total FI charge is concentrated in a layer
{\it less} than the C$_{60}$ diameter (0.7nm) and the center of gravity is 
strongly
shifted towards the interface. Our approach
 supposes the variation of the potential to be small over molecular
dimension, therefore it is valid for fields up 
to 2$\times 10^{6}$V/cm
corresponding to $n_s$=3.3$\times 10^{12}$cm$^{-2}$, where the system is 
already degenerated.
For smaller distances the approximation formally does not work 
(but similar ones give even for metal surfaces with a strongly
 varying
potential reasonable results). 
Note that  within a 
microscopic approach Wehrli et al.\ \cite{wehrli}
 came to the same conclusion:  
the FI charge is indeed confined to one C$_{60}$ layer, only. However,
the questions of the detrimental role 
of fluctuations should be addressed since they 
might destroy any FISC in a truly 2D-system
(see e.g.\ the lack of superconductivity in an
isolated CuO$_2$ bilayer\cite{kim}).

Next, we consider within the LDA-FPLO scheme\cite{koepernik99} 
the electronic structure for 3D undoped anthracene. 
 We calculated its band structure  in the monoclinic
spacegroup P21/c (No.\ 14) with the lattice constants $a$ = 11.172 \AA
, $b$ = 6.0158 \AA ,  $c$ = 8.553 \AA , and the monoclinic angle $\beta$
= 124.596$^\circ$ (see Fig.\ 1).
 Also for anthracene 
 the LDA underestimates  as usual the  gap value:  2.6 eV 
vs.\ 4.1 eV 
(exp.). 
\noindent
The band 
width $\sim$ 150 meV (see Fig.\ 2) 
 is rather small compared with the expected value of the 
 Hubbard $U\sim$ 2 to 3 eV ($\approx$ 1.6 eV for C$_{60}$ \cite{lof} which 
 gives a lower 
 bound for the less screened smaller PA molecules). 
 Then, formally, the injected charge carriers 
 in the FI effective 2D PA 
 surface band  at half-filling would be even stronger correlated than those 
 in undoped
 cuprates.  So, an antiferromagnetic (afm) Mott-Hubbard insulator might be 
 expected
 at $T=0$. However,    
 from the non-$\cos$ shape (i.e.\ beyond the n.n.\ tight-binding picture) 
  of the 
 dispersion along $b$ the presence of strong "frustrations"
 (for a corresponding afm $S$=1/2 Heisenberg-model) can be derived. 
 This  might be  important in explaining the metallicity 
 at  half-filling where the maximal $T_c$ is observed. To illustrate this 
 point,  we consider a 1D example (see Fig.\ 3)
 where the optical gap derived from the 
 optical conductivity (calculated using standard Lanzcos and continued fraction 
 methodes) is shown. For strong $t_2$ 
 the Mott-Hubbard (charge) gap could be strongly reduced and/or a spin gap
 \cite{daul},
  helpful for FISC, might occur.
  \begin{figure}[b!]
\begin{center}
\includegraphics[scale=0.3]{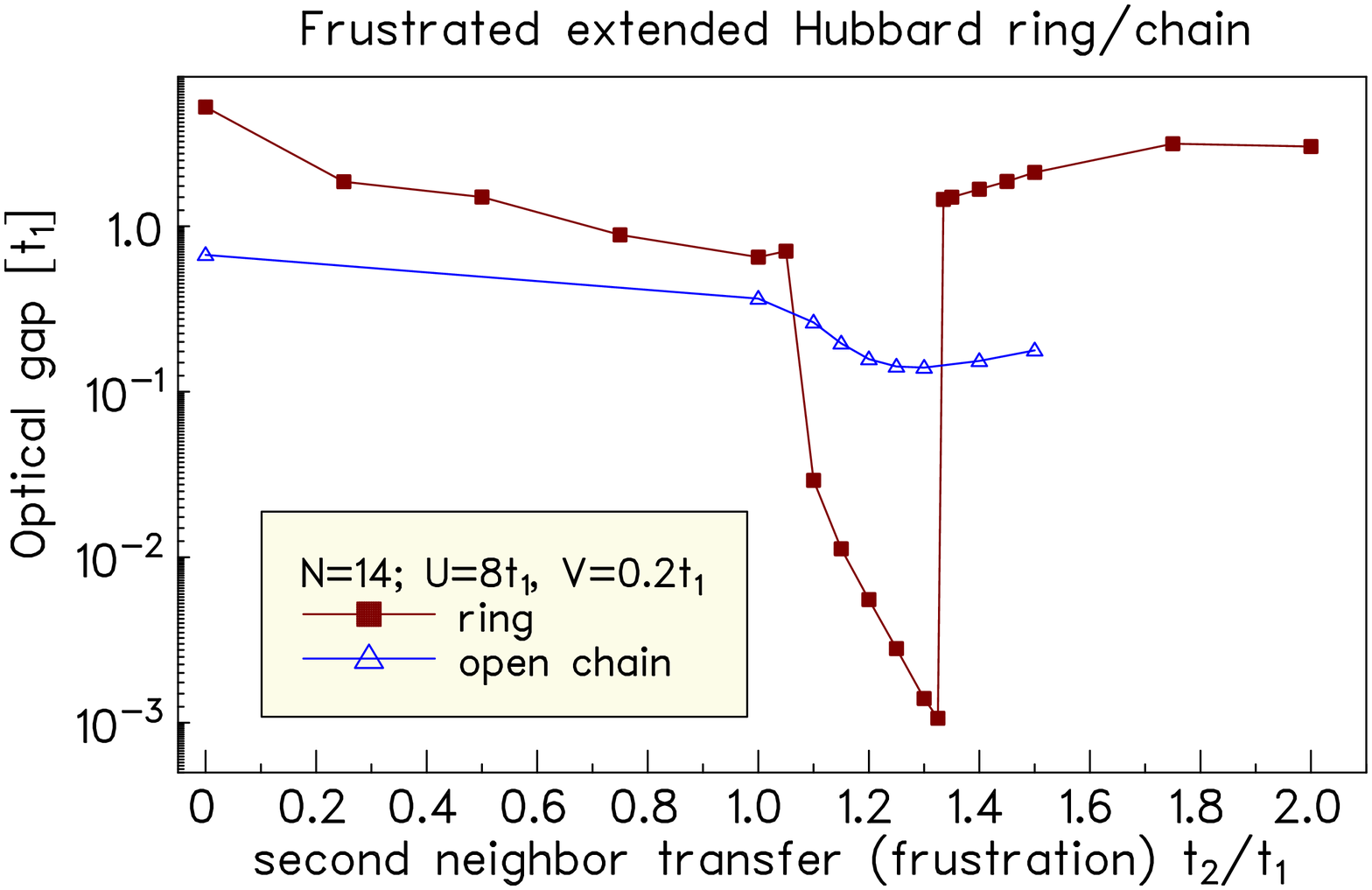}
\includegraphics[scale=0.3]{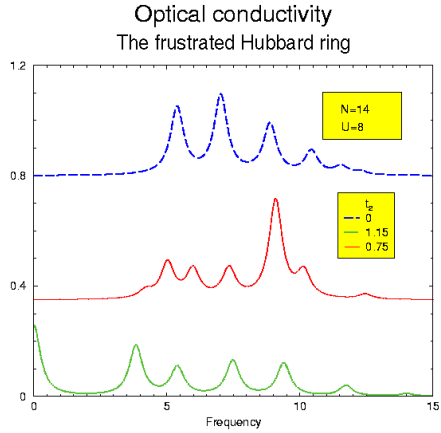}
\caption{ Upper panel: Optical gap vs.\ second neighbor transfer integral $t_2$ 
(in units of the nearest neighbor transfer integral $t_1$) for  strongly 
correlated chains 
 with $N=14$ sites and weak intersite Coulomb interaction $V$
at half filling; Lower panel: optical conductivity (in arbitrary units) vs.\ 
frequency (in units of $t_1$).}
\end{center}
\end{figure}
 The inspection of the optical conductivity shows that with increasing $t_2$ 
 the spectral feature attributed to the charge gap remains almost unshifted
 whereas some subgap absorption occurs which tends to a Drude-like peak in the 
 metal-like state.
 
 A phenomenological Eliashberg analysis has been 
 considered by us in Ref.\ \onlinecite{drechsler02}. 
 Such an approach would be supported 
by the direct measurement  of the
Eliashberg function $\alpha^2F(\omega)$ derived from the tunnel current 
such as reported for pentacene\cite{batlogg01}. 
From the calculated density of states (DOS) for the  
 conduction band the 
el-ph coupling constant 
$\lambda=N(0)V_{ph}\approx N(0)1800\mbox{meV}/N_\pi \approx 1.76$ was
    estimated 
  using $V_{ph}$ of Ref.\
\onlinecite{devos98}. Finally, from $\lambda$ and the  
reported $T_c$ the value of the Coulomb pseudopotential 
$\mu^*\approx 0.59$ was determined 
within   Eliashberg theory.
 The reported different $T_c$-values of PA and C$_{60}$
are affected by the value of the   
 Coulomb pseudopotential $\mu^*$. In particular,  
anthracene shows an enhanced $\mu^*$-value which
is ascribed to reduced screening on its small molecules. This is 
considered as the main reason for the  low  $T_c$ 
despite  the  strong el-ph coupling.
 Adopting the same DOS and an extrapolated 
 nearly saturated value
$\mu^*\approx 0.61$ like in anthracene, 
for naphtalene (the next shorter PA) 
even  
$T_c\sim$ 30 K  would be expected, while 
with $\mu^*\approx 0.45$ for the longer 
hexacene   $ T_c \sim$
0.6 K should almost vanish.

To summarize, the charge carriers induced by very strong electric 
 fields $\sim 10^{8}$V/cm in PA and C$_{60}$ are confined to one monolayer.
  In the case of PA they should be described by nontrivial 
 2D tight-binding models with possible  new many-body effects. 
 We believe that these results remain valid independently 
 of the reliability of the reported  
 data\cite{schoen00a,schoen00c,batlogg01}.
 On the contrary, 
 our 
 FISC analysis does rest on them. Hence, any confirmation, modification, or 
 even 
 refutation at all
 by other experimental groups would be highly desirable.
 
 We thank the Deutsche Forschungsgemeinschaft and the DAAD (H.R.) 
 for financial support under project 
 Es85/8-1.

\end{document}